\documentclass[aoas,MSNbibl,nameyear,rotating,dvips]{arximspdf}
\usepackage{dcolumn}
\usepackage{graphicx}
\usepackage{url,breakurl}

\doi{10.1214/14-AOAS768} 
\volume{8}
\issue{4}
\pubyear{2014}
\firstpage{2122}
\lastpage{2149}
\docsubty{FLA}

\makeatletter
\newcolumntype{d}[1]{D{.}{.}{#1
}}
\makeatother

\begin{document}
\begin{frontmatter}

\title{Global estimation of child mortality
using a~Bayesian B-spline Bias-reduction model\thanksref{T1}}
\runtitle{Global estimation of child mortality}
\thankstext{T1}{Supported by Grants R-155-000-099-133 and R-155-000-146-112
at the National University of Singapore and the United Nations
Children's Fund. }

\begin{aug}
\author{\fnms{Leontine} \snm{Alkema}\corref{}\ead[label=e1]{alkema@nus.edu.sg}}
\and
\author{\fnms{Jin Rou} \snm{New}\ead[label=e2]{jrnew@nus.edu.sg}}
\affiliation{National University of Singapore}
\runauthor{L. Alkema and J.~R. New}
\address{Department of Statistics and Applied Probability\\
National University of Singapore\\
Blk S16, Level 7, 6\\
Science Drive 2\\
Singapore 117546\\
Singapore\\
\printead{e1}\\
\phantom{E-mail:\ }\printead*{e2}} 
\end{aug}

\received{\smonth{5} \syear{2014}}

%
\begin{abstract}
Estimates of the under-five mortality rate (U5MR) are used to track
progress in reducing child mortality and to evaluate countries'
performance related to Millennium Development Goal 4. However, for the
great majority of developing countries without well-functioning vital
registration systems, estimating the U5MR is challenging due to limited
data availability and data quality issues.

We describe a Bayesian penalized B-spline regression model for
assessing levels and trends in the U5MR for all countries in the world,
whereby biases in data series are estimated through the inclusion of a
multilevel model to improve upon the limitations of current methods.
B-spline smoothing parameters are also estimated through a multilevel
model. Improved spline extrapolations are obtained through logarithmic
pooling of the posterior predictive distribution of country-specific
changes in spline coefficients with observed changes on the global
level.

The proposed model is able to flexibly capture changes in U5MR over
time, gives point estimates and credible intervals reflecting potential
biases in data series and performs reasonably well in out-of-sample
validation exercises. It has been accepted by the United Nations
Inter-agency Group for Child Mortality Estimation to generate estimates
for all member countries.
\end{abstract}

%
\end{frontmatter}

\section{Introduction}\label{sec1}
The under-five mortality rate (U5MR) is a key barometer of the
well-being of a country's children and, more broadly, an indicator of
socioeconomic progress. The U5MR is strictly not a rate, but the
probability that a child born in a given year will die before reaching
the age of five if subject to current age-specific mortality rates (UN
IGME 2013),\nocite{IGME2013} often expressed as the number of deaths
per 1000 live births. National estimates of the U5MR are used to track
progress in reducing child mortality and to evaluate countries'
performance with respect to the United Nations' Millennium Development
Goal 4 (MDG 4), which calls for a two-thirds reduction in the U5MR
between 1990 and 2015 (UN IGME 2013), \nocite{IGME2013} corresponding
to an annual rate of reduction of 4.4\%.

For the great majority of developing countries without well-functioning
vital registration systems, estimating levels and trends in U5MR is
challenging, not only because of limited data availability but also
because of issues with data quality. Every year, the United Nations
Inter-agency Group for Child Mortality Estimation (UN IGME, including
the United Nations Children's Fund, the World Health Organization, the
World Bank, and the United Nations Population Division) produces and
publishes estimates of child mortality comparable across countries and
years for 194 countries. In 2012, a Loess regression model was used to
estimate the U5MR (UN IGME 2012). \nocite{IGME2012}
For each country, the default setting for its smoothness parameter
$\alpha$ was determined by the type and availability of data in the
country. A bootstrap method was used to assess the uncertainty in the
U5MR estimates [\citet{AlkemaNew2012}]. A number of limitations with this
approach were identified. The first limitation was that for a subset of
countries, the fitted Loess curve was deemed to not fit the data well
and post-hoc adjustments in the $\alpha$ value were necessary. The
second limitation was that all observations were weighted equally to
obtain point estimates; standard errors, potential data biases and
indicators of data quality were not accounted for. The calibration of
the resulting point estimates and uncertainty intervals left room for
improvement.

Alternative methods for estimating child mortality for all countries
have been developed by the Institute for Health Metrics and Evaluation
(IHME) [\citet{Rajaratnametal2010,Wangetal2012}], which uses Gaussian
process regression modeling to obtain U5MR estimates. A model
validation exercise to check model performance based on the 2010
version of the IHME approach also indicated room for improvement [\citet
{Alkemaetal2012a}], possibly explained by the approach not fully
accounting for potential data biases. To the best of our knowledge, the
same exercise has not been repeated for the most recent iteration of
the IHME model [\citet{Wangetal2012}]. We expect that issues with model
calibration have not yet been fully addressed given that the data model
has not been updated to incorporate the possibility of data biases.

In this paper we propose an alternative U5MR estimation approach to
improve upon the limitations and lack of calibration of existing
methods. The approach is given by a Bayesian B-spline Bias-reduction
model, referred to as the B3 model. The UN IGME has decided to use the
B3 model to assess countries' progress toward MDG 4 and B3 estimates
are included in ``A~Promise Renewed Progress Report 2013'' [\citet
{APR2013}] and the ``Child Mortality Report 2013'' (UN IGME 2013).
\nocite{IGME2013}

The paper is organized as follows. Section~\ref{sec-background} provides background
information on child mortality estimation. In Section~\ref{sec-overview} we present the
B3 model specification, followed by validation results and resulting
U5MR estimates in Section~\ref{sec4}. We end with a discussion of the model and
scope for future research.

\begin{figure}

\includegraphics{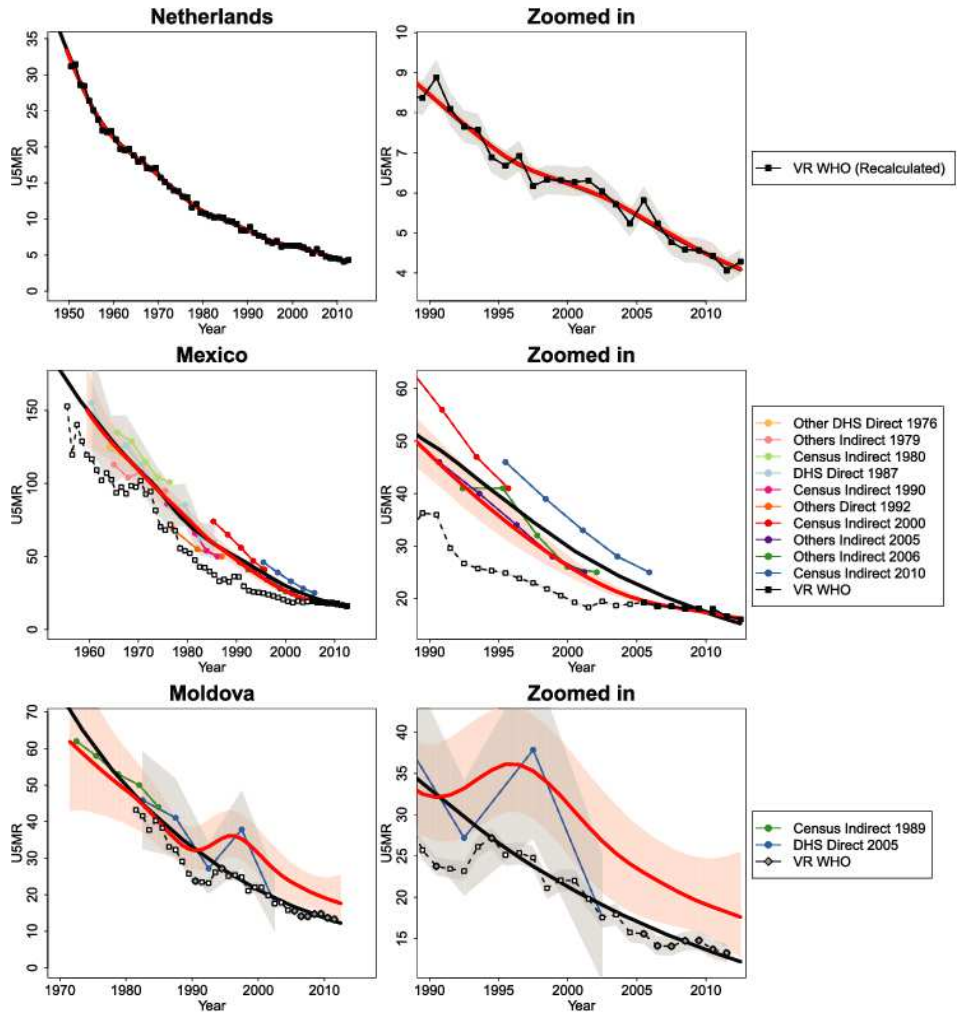}

\caption{U5MR data series and estimates for the Netherlands,
Mexico and Moldova.
Connected dots represent data series from the same source, as explained
in the legend.
B3 estimates are illustrated by the solid red lines and 90\% CIs are
shown by the red shaded areas.
The fitted Loess curve based on UN IGME 2012 methodology is illustrated
with the solid black line.
Shaded areas around series of observations represent the sampling
variability in the series
(quantified by two times the sampling standard errors).} \label{fig-countries}
\end{figure}

\begin{figure}

\includegraphics{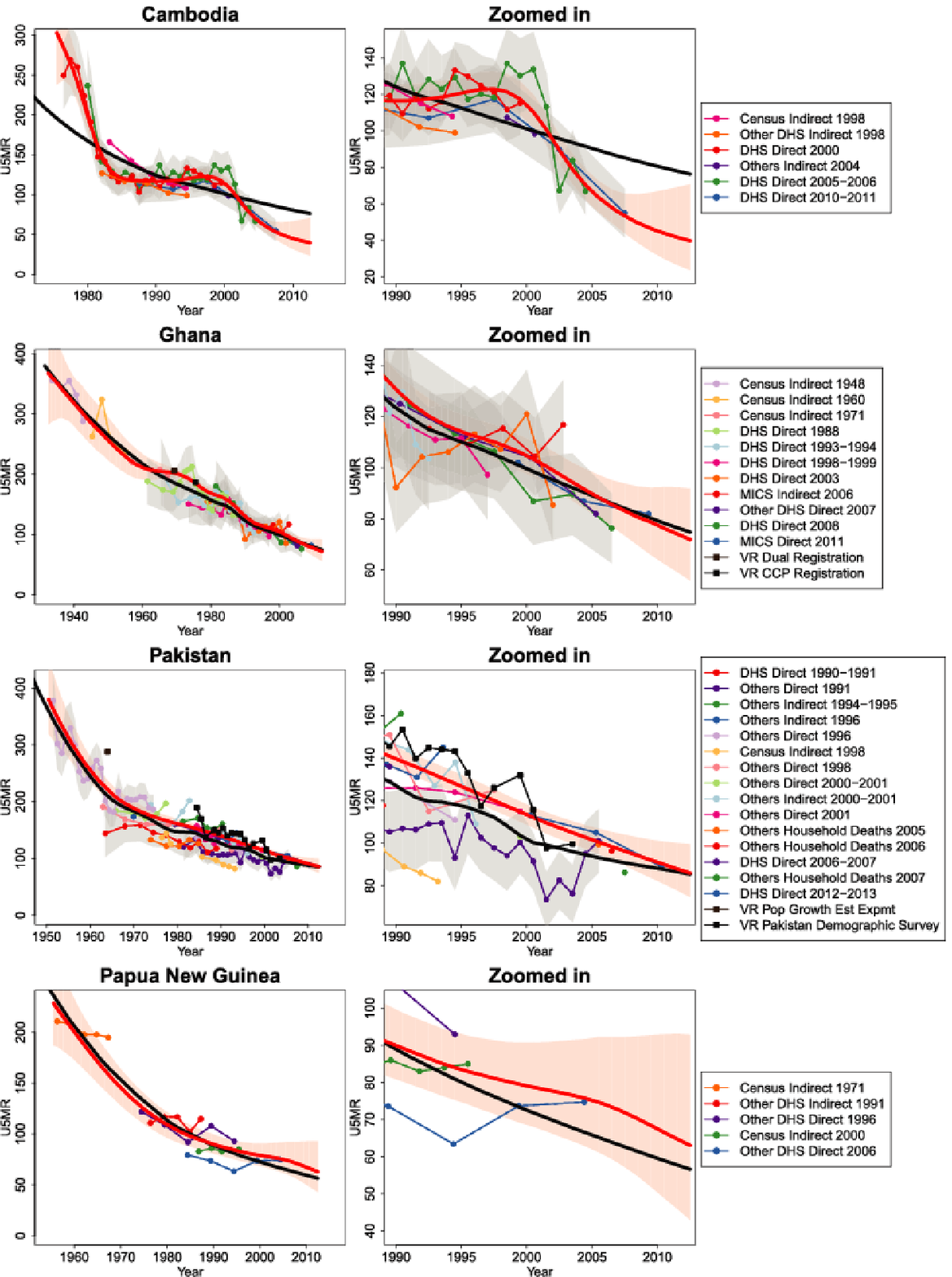}

\caption{U5MR data series and estimates for Cambodia, Ghana,
Pakistan and Papua New Guinea.
Connected dots represent data series from the same source, as explained
in the legend.
B3 estimates are illustrated by the solid red lines and 90\% CIs are
shown by the red shaded areas.
The fitted Loess curve based on UN IGME 2012 methodology is illustrated
with the solid black line.
Shaded areas around series of observations represent the sampling variability
in the series (quantified by two times the sampling standard errors).}
\label{fig-countries2}
\end{figure}

\section{Background}\label{sec-background}
U5MR data series are constructed from information from vital
registration (VR) and sample vital registration (SVR) systems, surveys
and censuses. U5MR data for selected countries are shown in Figures~\ref{fig-countries} and~\ref{fig-countries2}. The selected countries differ
with respect to U5MR level and trend, as well as data availability and
data quality.

In the Netherlands, data from the VR system capturing all births and
deaths are available since 1940. Such data from well-functioning VR
systems are the preferred data source for calculating U5MR. However, in
2013, 60 countries for which the UN IGME produces U5MR estimates did
not have any data from VR systems. Among the 135 countries with VR or
SVR systems, recording of birth and/or deaths is not necessarily
complete; illustrations are given for Mexico and Moldova. In Mexico, VR
data were deemed complete only since 2005. For Moldova, VR data are
considered incomplete for all observation years.

For countries without (or with limited information from)
well-functioning VR systems, complete or summary birth histories of
women, collected in surveys and censuses, are often the main source of
information on U5MR. A complete birth history lists all the live births
a woman has had, including information on the date of birth of each
child, whether the child is still alive, and if the child has died, the
age at death. U5MR observations are calculated from such information
through a synthetic cohort approach, whereby for a given period before
the survey, survival probabilities are calculated for small age
intervals and combined to obtain the U5MR for that period [\citet
{PedersenLiu2012}]. These observations are referred to as direct
estimates of U5MR. Many of these direct series are obtained from
complete birth histories that were collected as part of the
international household survey program Demographic and Health Surveys
(DHS). Other direct series are obtained from data from survey programs
similar to the DHS [here referred to as Other DHS as opposed to
(Standard) DHS], as well as other national surveys (referred to as
Others Direct). Examples of direct series are shown in Figures~\ref{fig-countries} and~\ref{fig-countries2}. Because of the retrospective
nature of the data, direct series can extend for up to decades before
the survey. For example, the DHS in Cambodia that was carried out in
2005--2006 provides data from 1979 to 2004.

As the name suggests, summary birth histories provide a summary of
complete birth histories: they list the number of live births a woman
has had and the number of children that have died. These summarized
histories are more commonly collected than complete birth histories
because of the simplicity of data collection. For summary birth
histories, demographic models are used to calculate the U5MR from the
recorded proportion of dead children for different time references
[\citet{brass1964,un1983}]. Because of the dependency on models, these
estimates based on summary birth histories are referred to as indirect
estimates. Indirect series are most commonly obtained using information
from censuses and surveys such as the Multiple Indicator Cluster Survey
(MICS), an international survey program that collects summary birth
histories in many developing countries. Examples of indirect series are
shown in Figures~\ref{fig-countries} and~\ref{fig-countries2}. As
discussed for direct data series, indirect series also provide data
points for a long retrospective period. For example, the Cambodian
census from 1998 provides indirect estimates from 1983 to 1994.

The availability of nationally-representative surveys and censuses
carried out in developing countries varies greatly. For instance, a
large number of data series are available from various sources in
Pakistan, but only five data series are available for Papua New Guinea.
Moreover, data series do not necessarily tell a similar story about
levels and/or trends in U5MR. For example, in Papua New Guinea, there
are large differences between U5MR estimates from the various sources.
In Pakistan, the DHS 2006--2007 survey suggests lower levels of U5MR
than data from its sample registration system. The spread in data
points for countries without data from well-functioning VR systems is
not specific to the selected countries in Figures~\ref{fig-countries}
and~\ref{fig-countries2}, but is observed in many developing countries,
as U5MR data are associated with a variety of data quality issues.
Apart from sampling error, observations from non-VR sources may also be
subject to bias and nonsampling error, for example, because of recall
biases when collecting birth histories. Specific data series may be
entirely biased upward or downward, for example, based on inaccuracies
in the indirect estimation method that was used to translate the
summary birth histories from a census or survey in U5MR observations.

Given issues with data quantity and quality, estimating the U5MR is
challenging for many countries. A modeling approach needs to be
flexible enough to capture short-term fluctuations in U5MR without
being overly sensitive to erroneous data fluctuations.

\begin{sidewaysfigure}

\includegraphics{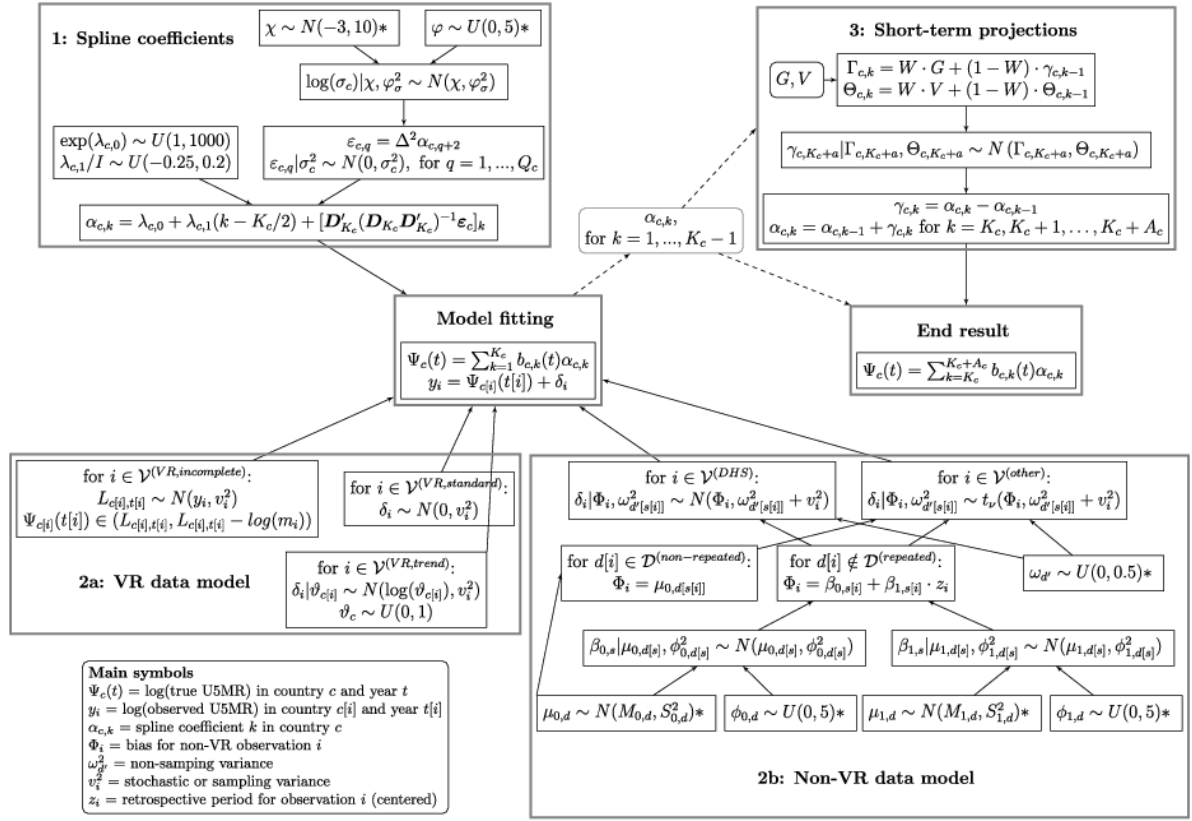}
\vspace*{-8pt}
\caption{Model overview. This chart summarizes the model used
to estimate the U5MR. In the center
is the description of the ``Model fitting'' part, where $\Psi_{c}(t)$
refers to the true U5MR on the log-scale,
which was modeled with a Bayesian penalized spline regression model, as
summarized in block 1
(see Section~\protect\ref{sec-methods-model}). The models for the error
term $\delta_i$ for observed log(U5MR)
are described separately for VR and non-VR data in blocks 2a and 2b
(see Section~\protect\ref{sec-methods-datamodel}).
Short-term projections are summarized in block 3 (see Section~\protect\ref{sec-project}).}
\label{fig-overview}
\end{sidewaysfigure}

\section{Constructing U5MR estimates}\label{sec-overview}
We developed a modeling approach that combines a flexible curve fitting
method with a comprehensive data model to account for data quality
issues. In the model description, lowercase Greek letters refer to
unknown parameters, uppercase Greek letters to functions of unknown
parameters, and Roman letters to fixed variables, including data
(lowercase). $\Lambda_{c}(t)$ denotes the quantity of interest, the
true U5MR in country $c$ in year $t$. U5MR observations are combined
across countries and indexed by $i = 1, 2,\ldots, N$; $u_i$~denotes
observed U5MR for observation $i$ in country $c[i]$ and year $t[i]$.

The complete model overview is given in Figure~\ref{fig-overview}. In
the center of the overview and the model is the description of the
``Model fitting'' for the true U5MR on the log-scale, $\Psi_{c}(t) =
\log(\Lambda_c(t))$ for country $c$ at time $t$. log(U5MR) was modeled
with a Bayesian penalized spline regression model, explained further in
Section~\ref{sec-methods-model} and summarized in block 1 (spline coefficients) of Figure~\ref{fig-overview}. For U5MR observations, we assumed
%
\begin{equation}
\label{eq-main} y_i = \Psi_{c[i]}\bigl(t[i]\bigr) +
\delta_i,
\end{equation}
where $y_i = \log(u_i)$ and $\delta_i$ is the error term on the log-scale.
The data and specification of error term $\delta_i$ are discussed
further in Section~\ref{sec-methods-datamodel} and summarized in blocks~2a and b (VR and non-VR data model) in Figure~\ref{fig-overview}.
Finally, short-term projections are discussed in
Section~\ref{sec-project} and summarized in block 3 (short-term projections).

Our analysis included 194 countries. For countries with high HIV
prevalence, conflicts or natural disasters,
we applied a modified estimation method based on the UN IGME 2012
estimation method, as explained in \citet{AlkemaNew2013}.

\subsection{Bayesian penalized spline regression}\label{sec-methods-model}
The regression spline model for log-transformed U5MR, $\Psi_{c}(t)$ in
equation (\ref{eq-main}), is given by
%
\begin{equation}
\label{eq-spline} \Psi_c(t) = \sum_{k=1}^{K_c}
b_{c,k}(t)\alpha_{c,k},
\end{equation}
where $\alpha_{c,k}$ refers to spline coefficient $k$ in country $c$
and $b_{c,k}(t)$ the $k$th B-spline in country $c$, evaluated in year
$t$. $K_c$ refers to the index of the most recent spline which is
nonzero during the observation period.

In this application, B-splines, as discussed in
\citeauthor{Eilers1996} (\citeyear{Eilers1996,Eilers2010}), were used, specifically third-degree (cubic)
B-splines, illustrated for selected countries in (the bottom of)
Figure~\ref{fig-ppdsimu}. Equally spaced knots were used such that the
resulting splines are nonzero for a total of $4\cdot I$ years, where
$I$ refers to the in-between-knots interval length. The same interval
length of 2.5 years was used in each country regardless of the
number/spacing of observations, to be able to exchange information
across countries about the variability in changes between spline
coefficients and assess the uncertainty in periods with limited data
(further explained below).

When fitting the spline model from equation (\ref{eq-spline}) to the
observations, second-order differences in adjacent spline coefficients
($\Delta^2\alpha_k = \alpha_k - 2\alpha_{k-1} + \alpha_{k-2}$) are
penalized to guarantee smoothness of the resulting U5MR trajectory. To
implement the smoothing, for each country $c$, spline coefficients
$\alpha_{c,k}$ for $k=1,2,\ldots, K_c$ were rewritten as follows
[\citet{Currie2002,Eilers1999,Eilers2010}]:
%
\begin{equation}
\label{eq-mm} \alpha_{c,k} = \lambda_{c,0} +
\lambda_{c,1}(k - K_c/2)+\bigl[\mathbf {D}_{K_c}'
\bigl(\mathbf{D}_{K_c}\mathbf{D}_{K_c}'
\bigr)^{-1}\bolds{\varepsilon}_c\bigr]_k,
\end{equation}
where $\lambda_{c,0}$ and $\lambda_{c,1}$ are the unknown level and
slope parameters for the spline coefficients in country $c$ and
parameter vector $\bolds{\varepsilon}_c = (\varepsilon_{c,1},\ldots,
\varepsilon_{c,Q_c})'$ contains the $Q_c = K_c-2$ second-order
differences in the spline coefficients, $\varepsilon_{c,q} = \Delta
^2\alpha_{c,q+2}$ for $q=1,\ldots, Q_c$; $[\mathbf{D}_{K_c}'(\mathbf
{D}_{K_c}\mathbf{D}_{K_c}')^{-1}\bolds{\varepsilon}_c]_k$ refers to the $k$th
element of vector $\mathbf{D}_{K_c}'(\mathbf{D}_{K_c}\mathbf{D}_{K_c}')^{-1}\bolds
{\varepsilon}_c$, with known difference\vspace*{1pt} matrix $\mathbf{D}_{K_c}$ (defined
by $D_{K_c,i,i} = D_{K_c,i,i+2} = 1$, $D_{K_c,i,i+1} =-2$ and
$D_{K_c,i,j} = 0$ otherwise).

Second-order differences are penalized by imposing
%
\begin{equation}
\label{eq-us} \varepsilon_{c,q}|\sigma_c^2
\sim N\bigl(0, \sigma_c^2\bigr)\qquad\mbox{for } q = 1,
\ldots, Q_c,
\end{equation}
where variance $\sigma_c^2$ determines the extent of smoothing; a
smaller variance corresponds to smoother trajectories. In the limit
when $\sigma_c$ decreases to zero (as the penalty increases), a linear
fit for log(U5MR) is obtained.

The model was fitted in the Bayesian framework. When estimating the
spline coefficients, no information on levels or trends during the
observation period was exchanged across countries to avoid the
situation where estimates for a country~A with little information are
pooled downward because it is neighboring country~B, where much
progress has been made in reducing child mortality or vice versa.
Diffuse priors were used for the $\lambda_{c,0}$'s and the $\lambda
_{c,1}$'s (see Block 2 in Figure~\ref{fig-overview} and the \hyperref[app]{Appendix}).

Information on spline coefficients is exchanged across countries only\break 
through a multilevel model for the variance of the differences in the
spline coefficients, that is, the standard deviation of $\varepsilon_{c,q}$
was estimated hierarchically:
%
\begin{equation}
\label{eq-sigmac} \log(\sigma_c)|\chi,\varphi_{\sigma}^2
\sim N\bigl(\chi, \varphi_{\sigma}^2\bigr),
\end{equation}
where $\chi$ and $\varphi_{\sigma}^2$ refer to the mean and variance of
the log-transformed standard deviations. Given the limited information
on shorter term fluctuations in some countries, there was not
sufficient information to estimate the variance parameter for each
country separately. The hierarchical model allows for sharing of
information across countries about the variability in changes between
spline coefficients and for assessing the uncertainty in periods with
limited data. Diffuse prior distributions were assigned to $\chi$ and
$\varphi_{\sigma}^2$ (see the \hyperref[app]{Appendix}).

The in-between-knots interval length $I=2.5$ years was set by comparing
U5MR estimates obtained using a range of $I$'s, for the full data set as
well as for a subset of data in a validation exercise. U5MR estimates
were found to be similar for interval lengths up to around 3 years, but
for larger $I$, shorter term fluctuations were not captured, suggesting
that the intervals of up to 3 years can be used. Here $I=2.5$ years was
used such that each spline is nonzero for 10 years. At any time $t$,
there are four nonzero B-splines $b_{c,k}(t)$ such that $\sum_k
b_{c,k}(t) = 1$. In each country, knot placement was fixed by setting
$T_{c,K_c} = t_{n_c} + 1.5\cdot I$, where $t_{n_c}$ denotes the most
recent observation year and $T_{c,K_c}$ the knot for the $K_c$th spline
in country $c$ (motivated further in Section~\ref{sec-melding}).

\subsection{Database and data model}\label{sec-methods-datamodel}
Under-five mortality data for all countries were taken from the UN IGME
database. This database is publicly available on CME Info
(\surl{http://www.childmortality.org}).

Section~\ref{sec-background} provided an introduction to U5MR data
sources. A more detailed overview and explanation on data sources is
given elsewhere [\citet{Hilletal2012}]. The breakdown of the U5MR
observations by their main source types is given in Table~\ref{tab-database}. Based on potential differences in biases and
nonsampling errors across data sources (explained further below), a
distinction was made between series of observations from complete and
summary birth histories (direct and indirect estimates, resp.),
and observations based on different data sources and data collection
methods (e.g., VR systems, records based on household deaths and life
tables obtained from reports).

\begin{table}
\tabcolsep=0pt
\caption{Summary of the U5MR data series and observations in
the UN IGME 2013 database by source type} \label{tab-database}
\begin{tabular*}{\textwidth}{@{\extracolsep{\fill}}lcc@{}}
\hline
& \textbf{Number of} & \multicolumn{1}{c@{}}{\textbf{Number of}} \\
& \textbf{data series} & \multicolumn{1}{c@{}}{\textbf{observations}} \\
\hline
VR (including SVR) & 110 & 2968 \\
(Standard) DHS Direct (with reported sampling errors) & 203 & 2902 \\
(Standard) DHS Direct (without reported sampling errors) & \phantom{0}15 & \phantom{00}56 \\
Other DHS Direct (with reported sampling errors) & \phantom{0}49 & \phantom{0}634 \\
Other DHS Direct (without reported sampling errors) & \phantom{0}25 & \phantom{0}107 \\
MICS Indirect (with reported sampling errors) & \phantom{0}55 & \phantom{0}248 \\
MICS Indirect (without reported sampling errors) & \phantom{0}20 & \phantom{00}80 \\
Census Indirect & 228 & 1074 \\
Others Direct & 144 & \phantom{0}507 \\
Others Indirect & 168 & \phantom{0}793 \\
Others Household Deaths & \phantom{0}56 & \phantom{00}56 \\
Others Life Table & \phantom{0}56 & \phantom{00}56 \\
\hline
\end{tabular*}
\tabnotetext[]{}{Note: ``Other DHS'' refers
to nonstandard demographic and health surveys, that is, Special,
Interim and National DHS, Malaria Indicator Surveys, AIDS Indicator
Surveys and World Fertility Surveys.}
\end{table}

\subsubsection{Data model}
\textit{VR data from complete registration systems}.
The error distribution for observations from complete VR or SVR indexed
by $i \in\mathcal{V}^{(\mathrm{VR}\ \mathrm{standard})}$ is given by
\[
\delta_i \sim N\bigl(0, v_i^2\bigr),
\]
where $v_i$ is the stochastic error variance. The stochastic error
variance was calculated using a Poisson approximation and the delta
method, assuming that
\[
\label{eq-poissondist} D_{c, t} |\Lambda_{c, t} \sim \operatorname{Poisson}(B_{c, t}
\cdot\Lambda_{c, t}/1000),
\]
where $D_{c, t}$ is the number of under-five deaths and $B_{c, t}$ is
the number of live births for country $c$ in year $t$.

The number of births were obtained from the World Population Prospects
[\citet{UNWPP2010}] and stochastic errors were set to a minimum of 0.025
(i.e., 2.5\%). For VR-type data from sample vital registration systems
where the number of sampled live births was not available, it was set
to 0.1 (i.e., 10\%) based on the target standard error for the Indian
sample registration system (Census of India, 2011). \nocite{IndiaSRS}

VR observations were typically calculated for single-year periods but
longer periods were used for smaller countries in instances where the
coefficient of variation of the observation was larger than 10\% (due
to small numbers of births and deaths).

\textit{VR data from incomplete registration systems.}
For 10 countries in the regional grouping of the Central and Eastern
Europe/Commonwealth of Independent States (CEE/CIS) (namely, Armenia,
Azerbaijan, Georgia, Kazakhstan, Kyrgyzstan, Moldova, Tajikistan,
Turkmenistan, Ukraine and Uzbekistan), VR data were incomplete with
respect to the reporting of deaths (biased downward) and generally
excluded from the estimation procedure in previous rounds of UN IGME
estimation. However, although not informative about the level of U5MR,
these observations were deemed to provide information on U5MR in the
early 1990s and for recent years. During the early 1990s, in several
CEE/CIS countries, data from the VR suggested a plateauing of or even
an increase in U5MR. This is illustrated in Figure~\ref{fig-countries}
for Moldova. This observed trend is assumed to reflect a true
stagnation in progress in reducing U5MR. To use this information, we
incorporated the option to include incomplete VR data into the model to
inform trend estimates in the country-specific B3 model. We also
included the option to set upper and lower bounds for recent years.
(These options were used in the country-specific models, as described
in Section~\ref{sec-1country}.)

To use the observed trend in VR data in the early 1990s to inform the
U5MR estimates, the VR observation in 1990 and the maximum observed VR
observation from 1991 to 1995 in each CEE/CIS country were selected,
with indices denoted by index set $V^{(\mathrm{VR}, \mathrm{trend})}$. For each selected
observation $i \in V^{(\mathrm{VR}, \mathrm{trend})}$, the distribution of the error term
$\delta_i$ was given by
\begin{eqnarray*}
\delta_{i}|\vartheta_{c[i]} &\sim& N\bigl(\log(
\vartheta_{c[i]}), v_i^2\bigr),
\\
\vartheta_c &\sim& U(0,1),
\end{eqnarray*}
where country-specific bias parameter $\vartheta_{c[i]}$ was added such
that the two selected observations in country $c$ could inform the
trend in U5MR estimates but not the level.

For the most recent period starting from 2005, for a subset of CEE/CIS
countries, U5MR extrapolations based on the global model either
decreased below incomplete VR observations (where incomplete refers to
incomplete reporting of deaths resulting in downward biased VR
observations) or the extrapolation resulted in estimates far above VR
observations for which an external assessment of VR data by the UN IGME
suggested a minimum level of completeness ranging from 50\% to 90\%. We
resolved the U5MR discrepancies between the B3 extrapolations and
(assumed completeness of) VR data by including a subset of VR
observations as a minimum U5MR value into the model (accounting for
stochastic errors). More precisely, based on the most recent incomplete
VR observation $y_i$ (with $i \in\mathcal{V}^{(\mathrm{VR}, \mathrm{incomplete})}$), the
lower bound $L_{c[i],t[i]}$ for the log(U5MR) for country $c[i]$ in
year $t[i]$ was obtained as follows:
\[
L_{c[i],t[i]} \sim N\bigl(y_i, v_i^2
\bigr).
\]
For selected observations, where a minimum level of completeness $m_i$
was set for incomplete VR observation $y_i$, we also included the upper
bound $L_{c[i],t[i]} - \log(m_i)$ for log(U5MR). For example, if the
minimum completeness for observation $i$ is 80\%, then $m_i = 0.8$ and
the upper bound for the U5MR is given by $\exp(L_{c[i],t[i]})/m_i = \exp
(L_{c[i],t[i]})/0.8$. VR-based upper and lower bounds were incorporated
into the model by excluding any log(U5MR) estimates which fell outside
the interval $(L_{c,t}, U_{c,t})$.

\textit{Non-VR data}. For non-VR data, the data model needs to
account for (i) sampling and nonsampling errors, (ii) potential biases
in trends and levels of U5MR data series, and (iii) possibility of outliers.

For observations from Standard and Other DHS Direct series, indexed by
$i \in\mathcal{V}^{(\mathrm{DHS})}$, the error was assumed to be normally distributed
\[
\delta_i|\Phi_i,\Omega_i^2
\sim N\bigl(\Phi_i,\Omega_i^2\bigr),
\]
with mean bias $\Phi_i$ and standard deviation $\Omega_i$. For
observations from other source types, indexed by $i \in\mathcal
{V}^{(\mathrm{other})}$, posterior predictive checks suggested that more
outliers were present, therefore, a $t$-distribution with unknown $\nu$
degrees of freedoms was used:
\begin{eqnarray*}
\delta_i|\Phi_i, \Omega_i^2
&\sim& t_\nu\bigl(\Phi_i, \Omega_i^2
\bigr),
\\
\nu&\sim& U(2,30),
\end{eqnarray*}
where $t_\nu(\Phi_i, \Omega_i^2)$ denoted a $t$-distribution with $\nu$
degrees of freedom, centered at $\Phi_i$ and rescaled by $\Omega_i$.

For observations from non-VR source types $d$ with potentially multiple
observations per series, mean biases were modeled as a linear function
of the retrospective period of the observation in the survey (the
difference between the observation reference date and the date of the
survey/census). This setup was motivated by known problems with
retrospective data, such as the occurrence of recall biases and
violations of modeling assumptions when calculating indirect U5MR
observations. The linear model for mean bias $\Phi_i$ for observation
$i$ is given by
\[
\Phi_i = \beta_{0,s[i]} + \beta_{1,s[i]}
\cdot z_i,
\]
where $\beta_{0,s[i]} + \beta_{1,s[i]}\cdot z_i$ represents the bias in
level and trend as a function of the retrospective period $z_i$ for
observation $i$ (centered at 10 years) in data series $s[i]$. The bias
in the level of the series $\beta_{0,s}$ was estimated with a
multilevel model:
\[
\beta_{0,s}|\mu_{0,d[s]}, \phi_{0,d[s]}^2
\sim N\bigl(\mu_{0,d[s]}, \phi _{0,d[s]}^2\bigr),
\]
where $d[s]$ refers to the source type of series $s$. The set of source
types with potentially multiple observations per series, indexed by
$\mathcal{D}^{(\mathrm{repeated})}$, is given by (Standard) DHS Direct, Other
DHS Direct (including Special, Interim and National DHS, Malaria
Indicator Surveys, AIDS Indicator Surveys and World Fertility Surveys),
MICS Indirect, Census Indirect, Others Direct and Others Indirect.
$\mu_{0,d}$ and $\phi_{0,d}^2$ represent source type-specific mean bias
and between-series variance, respectively. These two hyperparameters
were unknown and were assigned prior distributions, as illustrated in
Figure~\ref{fig-overview}.

A similar approach was used to estimate the slope $\beta_{1,s}$:
\[
\beta_{1,s}|\mu_{1,d[s]}, \phi_{1,d[s]}^2
\sim N\bigl(\mu_{1,d[s]}, \phi _{1,d[s]}^2\bigr),
\]
where $\mu_{1,d}$ and $\phi_{1,d}^2$ represent the mean slope and the
between-series variance for source type $d$. For observations
constructed from source types without repeated observations (reported
household deaths and reported life tables, $d \in\mathcal
{D}^{(\mathrm{nonrepeated})}$), we assumed that $\Phi_i = \mu_{0,d[s[i]]}$.

Scale parameter $\Omega_i$ was modeled as a combination of sampling
variance $v_i^2$ and nonsampling variance $\omega_{d'[s[i]]}^2$:
\[
\Omega_i^2= \omega_{d'[s[i]]}^2
+ v_i^2,
\]
where source type $d'[s]$ for series $s$ refers to a further breakdown
of source types to distinguish between DHS, Other DHS and MICS surveys
with and without reported sampling errors for their observations (as
indicated in Table~\ref{tab-database}). If the sampling standard errors
were not reported, a sampling standard error of 2.5\% was used for
Census Indirect observations and 10\% for all other observations.
Nonsampling variance refers to variability because of random errors
that 
arise through imperfections in the data collection process and is unknown.

Hyperparameters
$\mu_{0,d}, \phi_{0,d}^2, \mu_{1,d}, \phi_{1,d}^2$, $\omega_{d'}^2$ and
$\nu$ were assigned prior distributions, as listed in the \hyperref[app]{Appendix}.
Diffuse priors were used for all hyperparameters, with the exception of
the mean bias $\mu_{0,d}$ for the DHS Direct series: an informative
prior distribution was used, based on an analysis of these biases in
the previous 2012 round of UN IGME estimates.

\subsection{Extrapolation using a logarithmic pooling approach}\label
{sec-melding}\label{sec-project}
The one-step-ahead projection of a future change in spline coefficients
based on the penalized spline regression model is given by
%
\begin{equation}
\gamma_{c,k}|\gamma_{c,k-1},\sigma_c^2
\sim N \bigl(\gamma _{c,k-1},\sigma_c^2 \bigr),
\end{equation}
where $\gamma_{c,k}= \Delta\alpha_{c,k} = \alpha_{c,k} - \alpha_{c,k-1}$.
This extrapolation can result in a high probability of unusually low or
high projected rates of change in the spline coefficients for a
specific U5MR trajectory if $\sigma_c$ is large and/or if $\gamma
_{c,k-1}$ is unusually small or large. If projected changes in spline
coefficients are unusually low or high over longer periods, so are the
projected changes in the U5MR, potentially giving rise to unrealistic
U5MR projections. To overcome this potential problem with the spline
extrapolations, we implemented a logarithmic pooling procedure to
combine country-specific posterior predictive distributions (PPDs) for
changes in spline coefficients with a global PPD and verified whether
this approach improved out-of-sample projections. This procedure was
applied to modify the PPDs for $\alpha_{c,k}$ for $k=K_c, K_c+1,\ldots, P_c$, where $K_c$ and $P_c$ refer to the indices of the most recent
splines in the observation and projection periods, respectively.
Spline coefficient $\alpha_{c, K_c}$ was included in the set of
``projected'' coefficients to be pooled because it is based on very
limited information only; the $K_c$th spline is placed such that it is
nonzero only for 1.25 years during the observation period, from
$t_{n_c}-1.25$ to $t_{n_c}$.

The approach is summarized as follows (see also block 3 in Figure~\ref{fig-overview}): Let $\alpha_{c,k}^{(j)}$ denote the $j$th posterior
sample of spline coefficient $k$ for country $c$, $j=1,\ldots, J$ and
let $\Gamma_{c,k+1}^{(j)} = \Delta\alpha_{c,k+1}^{(j)} = \alpha
_{c,k+1}^{(j)} - \alpha_{c,k}^{(j)}$, the $j$th posterior sample of the
differences between two adjacent spline coefficients. After fitting the
B3 model, we obtain $\gamma_{c,k}^{(j)}$ for $k=1,2,\ldots, K_c-1$,
while $\gamma_{c,k}$'s for $k=K_c, K_c+1,\ldots, P_c$ are drawn from
a pooled PPD (see Figure~\ref{fig-ppdsimu}).
The pooled PPD is a combination of the ``model-induced''
country-specific PPD for $\gamma_{c,k}^{(j)}$, defined by the penalized
splines model and a global PPD for future changes in the spline
coefficients. The global PPD was based on the set of posterior median
estimates of the $\gamma_{c,k}^{(j)}$'s, $\hat{\gamma}_{c,k}$ for $c =
1, \ldots, C$ and $k= 2,\ldots, K_c-1$ (during the observation period
for each country). We used country-projection-step-specific logarithmic
pooling weights to obtain the same extent of pooling for all countries.
The resulting pooled PPD for $\gamma_{c,K_c+a}^{(j)}$ for $a \geq0$ is
given by
\[
\gamma_{c,K_c+a}^{(j)}|\Gamma_{c,K_c+a}^{(j)},
\Theta_{c,K_c+a}^{(j)} \sim N \bigl(\Gamma_{c,K_c+a}^{(j)},
\Theta_{c,K_c+a}^{(j)} \bigr),
\]
where
\begin{eqnarray*}
\Gamma_{c,k}^{(j)} &=&W\cdot G+ (1-W)\cdot
\gamma_{c,k-1}^{(j)},
\\
\Theta_{c,k}^{(j)} &=& W\cdot V + (1-W)\cdot
\Theta_{c,k-1}^{(j)},
\end{eqnarray*}
with $G$ and $V$ equal to the median and variance of the $\hat{\gamma
}_{c,k}$'s, respectively, $\gamma_{c,K_c-1}^{(j)} = \alpha
_{c,K_c-1}^{(j)} - \alpha_{c,K_c-2}^{(j)}$ and $\Theta_{c,K_c-1}^{(j)}
= \sigma_c^{(j)}$.\vspace*{1pt}

The overall pooling weight $0\leq W \leq1$ was chosen through an
out-of-sample validation exercise (described in Section~\ref{sec-methods-validation}). Further details of the logarithmic pooling
procedure are given in the \hyperref[app]{Appendix}.

\subsection{Computation}\label{sec-compu}
A Markov Chain Monte Carlo (MCMC) algorithm was employed to sample from
the posterior distribution of the parameters with the use of the
software \texttt{JAGS} [\citet{plummer2003}]. Six parallel chains with
different starting points were run with a total of 50,000 iterations in
each chain. Of these, the first 10,000 iterations in each chain were
discarded as burn-in and every 20th iteration after was retained. The
resulting chains contained 2000 samples each. Standard diagnostic
checks (using trace plots, the Raftery and Lewis diagnostic [\citeauthor{RafteryLewis1992a}
(\citeyear
{RafteryLewis1992a,RafteryLewis1996})] and the Gelman and Rubin
diagnostic [\citet{gelmanrubin1992}])
were used to check convergence.

Estimates of relevant quantities are given by the posterior medians,
while 90\% credible intervals (CIs) were constructed from the 5\% and
95\% percentiles of the posterior sample. Given the inherent
uncertainty in U5MR estimates, 90\% CIs are used by UN IGME instead of
the more conventional 95\% ones.

\subsection{Country-specific UN IGME model and adjustments}\label
{sec-igmemodel}\label{sec-1country}
The B3 model was accepted by UN IGME to evaluate countries' progress
and performance in reducing U5MR. For this purpose, a computationally
cheaper and more user-friendly country-specific model was implemented,
with noncountry-\break specific parameters (marked with a star in Figure~\ref{fig-overview}) fixed at the posterior medians from the global model
run, which resulted in very similar estimates. For the country-specific
runs, we ran 6 chains with a total of 35,000 iterations in each chain.
Of these, the first 10,000 iterations in each chain were discarded as
burn-in and every 20th iteration after was retained. The resulting
chains contained 1250 samples each.

After reviewing the estimates, two model adjustments were incorporated
in the country-specific models to consistently adjust the level of
under- or over-smoothing in a subset of countries [see \citet
{AlkemaNew2013} for details].

Adjustments were also applied to the Democratic Republic of Congo and
Somalia, where the U5MR data are not deemed to be representative of the
country's past. Specifically, B-splines corresponding to conflict
periods where the U5MR is unlikely to have declined were combined such
that only one spline coefficient was estimated for each conflict
period. The resulting fit is constant during the conflict periods.

\subsection{Model validation}\label{sec-methods-validation}
Model performance was assessed through an out-of-sample validation
exercise. Given the retrospective nature of U5MR data and the
occurrence of data in series, the training set was not constructed by
leaving out observations at random, but based on all available data in
some year in the past [\citet{Alkemaetal2012a}]; here 2006 was chosen. To
construct the training set, all data that were collected in or after
2006 were removed. For example, if a DHS was carried out in 2006, all
(retrospective) observations from that DHS were left out of the
training set. Fitting the B3 model to the training set resulted in
point estimates and CIs that would have been constructed in 2006 based
on the proposed method.

To validate model performance, we calculated various validation
measures based on the left-out observations and based on the estimates
obtained from the full data set and the estimates obtained from the
training data set. The validation measures considered were mean and
median errors, coverage of prediction intervals (to quantify the
calibration of the prediction intervals) and interval scores (to
quantify calibration and sharpness of the prediction intervals).

For the left-out observations, errors are defined as $e_i = u_i - \tilde
{u}_i$, where $\tilde{u}_i$ denotes the posterior median of the
predictive distribution for a left-out observation $u_i$ based on the
training set. Coverage is given by $1/N \sum1[u_i \geq
l_{c[i]}(t[i])] \cdot1[ u_i \leq r_{c[i]}(t[i])]$, where $N$ denotes
the total number of left-out observations considered and
$l_{c[i]}(t[i])$ and $r_{c[i]}(t[i])$ the lower and upper bounds of the
90\% predictions intervals for the $i$th observation. The (negatively
oriented) interval score $n_i$ for observation $i$ is given by \citet
{Gneiting2007}
\begin{eqnarray*}
n_i &=& \bigl(\log(r_{c[i]}) - \log(l_{c[i]})
\bigr) +2/x \bigl(\log (l_{c[i]}) - y_i \bigr)\cdot1
[u_i < l_{c[i]} ]
\\
&&{}+2/x \bigl(y_i-\log(r_{c[i]}) \bigr)\cdot1
[u_i > r_{c[i]} ],
\end{eqnarray*}
with significance level $x=0.1$. This score combines the width of the
prediction interval with a penalty for any intervals that do not
contain the left-out observation. The validation measures were
calculated for 100 sets of left-out observations, where each set
consisted of a random sample of one left-out observation per country.
Reported results include the median and standard deviation of the
validation measures based on the outcomes in the 100 sets.

``Updated'' estimates, denoted by $\widehat{\Lambda}_c(t)$ for country
$c$ in year $t$, refer to the median U5MR estimates obtained from the
full data set. The error in the estimate based on the training sample
is defined as $e_{c,t} = \widehat{\Lambda}_c(t)-\widetilde{\Lambda
}_c(t) $, where $\widetilde{\Lambda}_c(t)$ refers to the posterior
median estimate based on the training sample, while relative error is
defined as $e_{c,t}/\widehat{\Lambda}_c(t) \cdot100$. Coverage and
interval scores were calculated in a similar matter as for the left-out
observations, based on the lower and upper bound of the $90$\% CIs for
$\log$(U5MR) obtained from the training set.
Coverage, mean/median errors and interval scores were also evaluated
for the annual rate of reduction (ARR) from 1990 to 2005.

\begin{figure}[b]

\includegraphics{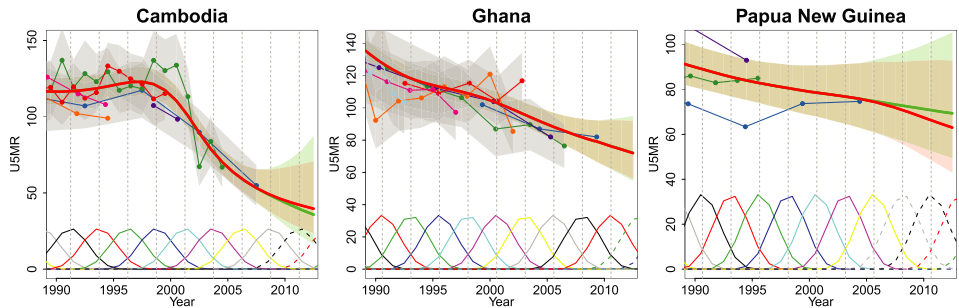}

\caption{Illustration of differences in estimates and
projections for Cambodia, Ghana and Papua New Guinea
between the unpooled (country-specific) and pooled
B-spline model projection approach. U5MR estimates
based on the unpooled (country-specific) projection
approach are displayed in green, and results based
on the pooling approach, using weight $W=0.5$, are
displayed in red. Solid lines denote posterior
medians and shaded areas denote 90\% CIs.
Additionally, the placement of B-splines
is shown in the bottom of the graph
(scaled vertically for display purposes)
in different colors. The dashed lines are
used for B-splines for which the corresponding
coefficient is included in the set of
``projected'' coefficients (with index $k \geq K_c$).
Gray dotted vertical lines indicate knot positions of the
B-splines.}\label{fig-ppdsimu}\label{fig-iluspool}
\end{figure}

Particular attention was paid to the performance of the B3 model for
the group of high mortality countries, where high here refers to a U5MR
of at least 40 deaths per 1000 births in 1990. This set was selected
because of the importance of the UN IGME U5MR estimates for tracking
progress in reducing child mortality. Crisis years and HIV adjustments
were not considered in the out-of-sample model validation because the
calculation of crisis-related and HIV-related U5MR is not included in
the B3 method (so it is not possible to reconstruct these estimates).

\section{Results}\label{sec4}
\subsection{Model validation and choice of pooling weight}
To set the pooling weight~$W$ (to combine the PPD of country-specific
changes in spline coefficients with the global PPD), validation
measures were obtained for $W = 0,0.1,\ldots, 0.6$, where $W=0$
corresponds to the ``no-pooling'' (country-specific) variant. In
general, differences between country-specific and pooling variants were
small for the posterior median point estimates but more noticeable for
projection intervals. An illustration of the default (unpooled) and
pooled projections (using $W = 0.5$) are shown in Figure~\ref{fig-iluspool} for Cambodia, Ghana and Papua New Guinea. The
introduction of the pooling procedure increased the U5MR projections in
Cambodia and led to a decrease in Ghana and Papua New Guinea, but
differences in point estimates were minor. Projection intervals varied
more across countries; the bounds were similar for weights 0 and 0.5
for Ghana, but narrowed down in Cambodia and were lower for the pooled
projections in Papua New Guinea.

Model validation results based on the left-out observations and the
comparison between estimates based on the training and full data set
are shown in Tables~\ref{tab-errors-ppd2}, \ref{tab-errors-ppd} and \ref
{tab-errors} in the \hyperref[app]{Appendix} for the range of pooling weights.
Differences in mean/median (absolute) errors were small. While for
median errors the comparison across the different weights varied by
indicator, mean errors generally decreased with increasing pooling
weights. Coverage and interval width scores for left-out observations
generally improved slightly with increasing pooling weight.
For the estimated U5MR and ARR, findings on coverage of 90\% credible
intervals were mixed, but mean interval scores for U5MR decreased with
increasing pooling weight.

Based on these findings, we chose to apply the pooling. Because
differences in validation outcomes were small when comparing the
results for $W=0.5$ to those with $W=0.6$, and because of the
convenient interpretation of $W = 0.5$ (the projected mean and variance
of the differences in the spline coefficients are the simple average of
the country-specific and global estimates), we set $W = 0.5$.
A~comparison of estimates and short-term projections based on $W = 0$ and
$W = 0.5$ for all countries is included in supplementary Figure S1
[\citet{AlkemaNew2014supp}].

With this choice of $W$, the model validation results for the B3 model
showed an improvement over those for the UN IGME 2012 estimation
approach. In a similar validation exercise carried out for the UN IGME
2012 estimation approach [\citet{AlkemaNew2012}], the updated estimate of
ARR for 1990--2005 (based on the full data set) was above the training
90\% CI for 16\% of the high mortality countries (11 out of 70
countries) and below that for only 6\% of those countries. This
indicates that declines in U5MR were underestimated for a substantial
proportion of high mortality countries. The same effect is observed in
the validation results for the B3 model but to a much lesser extent,
with only 9\% of the updated upper bounds for the ARR being too low and
3\% of the updated lower bounds being too high. Overall, the
calibration measures are better with the B3 model. Specifically, the
percentages of updated estimates falling below and above the 90\%
uncertainty intervals were 4\% and 5\%, respectively, for the U5MR in
2000 and 8\% and 1\% for the U5MR in 2005 in the B3 model. These
percentages were 10\% and 6\% for the U5MR in 2000 and 17\% and 7\% for
the U5MR in 2005 in the IGME 2012 estimation approach.

\subsection{Data model biases}\label{sec-res-datamodel}
\begin{figure}

\includegraphics{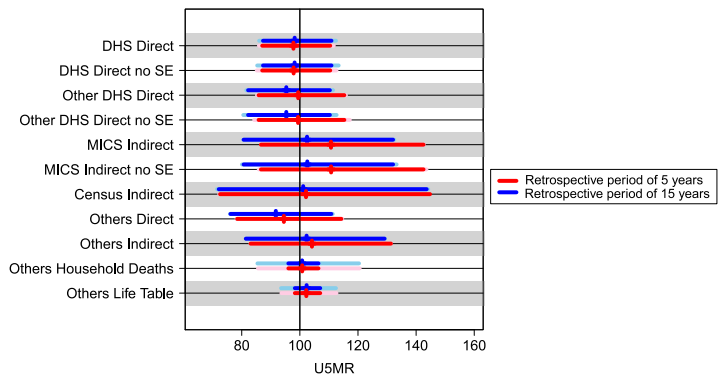}

\caption{Visualization of 90\% prediction intervals for new
data points by source type and retrospective period. For a ``true''
U5MR of 100 deaths per 1000 live births (represented by the black
line), the 90\% prediction interval for a U5MR observation with a
retrospective period of $5/15$ years is shown in light blue/pink
(excluding the sampling variability) and the predicted mean observed
U5MR is represented by the dark blue/red vertical line (the difference
between the mean U5MR and 100 represents the mean bias). The dark
blue/red horizontal line represents the 90\% prediction interval for an
observation based on uncertainty in the bias parameters only (excluding
sampling and nonsampling variability).}\label{fig-sourcetypePIs}
\end{figure}

Mean biases in U5MR levels and trends, as well as 90\% prediction
intervals for the expected range of U5MR values, were calculated based
on the posterior sample of data quality parameters and are visualized
in Figure~\ref{fig-sourcetypePIs} for the different source types. Mean
biases and prediction intervals are relative to the unknown true U5MR
level, which in the figure is assumed to be 100 deaths per 1000 live
births for ease of interpretation. The prediction intervals thus
illustrate the expected range of U5MR values for a ``new'' data series
when the true U5MR is 100 deaths per 1000 births. Results are shown
for retrospective periods of 5 and 15 years, thus for observations that
refer to 5 and 15 years before the survey date.

For indirect series, the 90\% prediction intervals based on uncertainty
in biases alone (the dark blue horizontal lines) are wide, indicating
substantial variability in biases across data series. For example, the
prediction interval ranges from about 87 to about 143 deaths per 1000
live births for an observation from a MICS Indirect series, with a
retrospective period of 5 years. The error variance tends to contribute
less to the width of the 90\% prediction intervals, implying that there
is significant variability in data series that is not attributed to
random error. For retrospective periods of 5 years, mean biases are
slightly positive for indirect series, but almost zero or negative for
direct series: observations from direct series tend to be below
indirect series for these retrospective periods.

\subsection{U5MR estimates}\label{sec-results}
U5MR data series and B3 estimates for all 194 countries are shown in
supplementary Figure S2 [\citet{AlkemaNew2014supp}]. For the selected
illustrative countries in Figures~\ref{fig-countries} and~\ref
{fig-countries2}, B3 estimates are displayed in the country-specific
figures, together with the estimates that would have been obtained
using the default Loess estimation approach used for constructing the
IGME 2012 estimates.

Point estimates from the B3 model and default Loess are almost
identical for the Netherlands during the entire observation period, but
differ for all or a subset of observation years in the other countries.
For Mexico, the trend in the Loess estimates for the late 2000s
contradicts the observed trend in VR data. B3 estimates take into
account the small stochastic error in the VR and follow the data points
closely. For Moldova, the inclusion of the VR observations in the early
1990s with a VR bias parameter for those years results in U5MR
estimates that capture the VR-indicated trend. The inclusion of VR data
for recent years guarantees that the point estimates and credible
intervals do not cross through the VR. In future revisions for Moldova,
a further extension could be to include all incomplete VR observations
as a minimum to avoid the situation in the early 1980s, when the lower
bound of the CI is below the incomplete VR.

For Ghana, B3 estimates and Loess estimates are similar. Small
differences are observed in the years with VR data, where the B3
estimates capture these points while the Loess does not. In more recent
years, the extrapolated decline is slightly steeper for the B3 model,
as indicated by the decline in the most recent observations.
Differences between B3 and Loess estimates are much larger in the other
countries in the figure. In Cambodia, the B3 estimates follow the trend
as observed in the data series, including the stagnation of child
mortality decline in the 1980s and 1990s and the more recent
acceleration in the decline of child mortality. The default Loess fit
does not capture these fluctuations. In the IGME 2012 method, this
country would be a candidate for an expert-based adjustment of the
Loess smoothing parameter to better capture the trend. In the B3
penalized spline model approach, such expert adjustments are not necessary.

In Pakistan, the B3 estimates follow the registration data. The DHS
from 2006--2007 does not bias downward the estimates (as observed in
the Loess estimates) because of the inclusion of bias parameters for
survey data; we estimate that the DHS Direct series is biased downward.
Last, in PNG, B3 estimates suggest a slightly flatter trend in U5MR
than the Loess during the 1980s and 1990s based on the lack of downward
trends in all individual series during that period.

\section{Discussion}\label{sec5}
The estimation of child mortality is challenging for the great majority
of developing countries without well-functioning VR systems due to
issues with data quantity and quality. In this paper, we described a
Bayesian penalized B-spline regression model to evaluate levels and
trends in the U5MR for all countries in the world. This model estimates
biases in data series for all non-VR source types using a multilevel
model to improve upon the limitations of current methods. Improved
spline extrapolations are obtained via logarithmic pooling of the
posterior predictive distribution of country-specific changes in spline
coefficients with observed changes on the global level. The proposed
model can flexibly capture changes in U5MR over time, provides point
estimates and credible intervals that take into consideration potential
biases in data series and gives better model validation results than
the UN IGME 2012 estimation approach.

The differences between the B3 estimates and the default Loess fits as
discussed in Section~\ref{sec-results} highlight the need for more
attention for appropriate data models in U5MR estimation. When treating
all observations equally, U5MR estimates can end up below (incomplete)
VR observations or follow a trend in U5MR that is dictated by the (lack
of) overlap of different data series with potentially different level biases.

While our data model overcomes the main limitations of the previous UN
IGME estimation methods, there remains room for improvement. The
primary issue with child mortality estimation is data quality. In the
B3 data model, we incorporated source-specific bias parameters, that
are drawn from a source type-specific distribution based on the
assumption that biases are comparable across data series of the same
source type. However, large variation exists across series; ideally,
external information on data quality should be included to distinguish
between the more or less reliable series in the database.
In a residual analysis, (absolute) residuals were plotted against a
number of data quality predictors (region that country belongs to,
series source type, series year, observation year, retrospective
period, level of U5MR in observation year, total fertility rate in the
series year and change in the total fertility rate in the last 15 years
before the series) to explore whether any of those covariates should be
incorporated into the model for biases in direct and indirect series.
Overall, the linear model without covariates seemed to work reasonably
well except for some DHS Direct series, for which an additional
negative bias for observations with retrospective periods shorter than
5 years may be present. This may be due to birth transference, whereby
dates of birth are incorrectly reported to avoid answering more
questions pertaining to those births in the DHS questionnaires [\citet
{Sullivan2008}]. Given the importance of the observations with short
retrospective periods in driving recent estimates and short-term
projections, this issue needs to be investigated more in future work.

Improved spline extrapolations were obtained via logarithmic pooling of
the posterior predictive distribution of country-specific changes in
spline coefficients with observed changes on the global level. While
short-term projections that are based only on country-specific
information may be preferred from a political/country-user point of
view, the pooling procedure was used because it was found to improve
out-of-sample model performance and deemed to lead to more plausible
projection intervals in countries where differences between the pooled
and unpooled predictions occurred. In summary, the pooling approach
reduces the probability of unrealistically high or low rates of changes
in extrapolations and also reduces the probability of sustained high or
low rates of change over longer projection periods by pooling the
predictive distribution for rates of change toward a global
distribution. This procedure did not result in large differences in
point estimates for the majority of countries (as illustrated in Figure
S1); its main effect was a reduction of upper bounds for the U5MR by
reducing the probability of very low or even negative rates of change.
Alternative projection methods may be considered, for example, based on
country-specific covariates which may be informative of U5MR declines.
However, given the limited availability of such covariates for recent
years, we did not pursue this research direction.

Ultimately, the issues of data quality and availability of more
recent data can only be resolved by implementing fully
functioning VR systems that can provide accurate data
on births and deaths in every country. However,
currently only about 50 countries have such VR
systems in place; the implementation of VR systems for all countries
remains an ambitious and long-term goal [\citet{RealTimeMeeting2012}]. In
the short term, the B3 model allows for inclusion of information from
incomplete VR systems, as illustrated for Moldova. The inclusion of
data from alternative data sources and the implementation of novel data
collection methods, that can provide accurate and timely child
mortality data [e.g., see \citet{Clarketal2008} and \citet
{Amouzou2011}], could further aid child mortality estimation. The
advantage of the use of the Bayesian framework in the B3 model is that
the model can be readily extended to incorporate such information into
the estimation process.

To assess progress toward MDG 4, much focus is placed on
the point estimates of the U5MR and ARR despite the large
uncertainty in estimates because communication of uncertainty in U5MR
estimates is challenging [\citet{oestergaard2013}]. To provide a
straightforward inclusion of the uncertainty assessment into the MDG 4
progress assessment, countries could be categorized by whether the
attainment of the MDG target of an ARR of 4.4\% is considered to be
unlikely, not clear or likely based on the uncertainty intervals of the
ARR estimate [\citet{AlkemaNew2012}].

Moving beyond the MDGs, the issue of inequality is likely to feature
prominently in the post-2015 development agenda. While the MDGs have
focused much attention on national, regional and global averages of key
indicators, they have also potentially masked growing disparities at
the intra-national level [\citet{UNTaskTeam2012}]. In light of this,
disaggregated estimates of child mortality (e.g., by state, wealth
quintile, residence) will be increasingly important to evaluate
progress for all population groups to better address inequalities.
Further work can be carried out to extend the B3 model so that this
growing body of disaggregated data can be fully utilized to produce
disaggregated estimates in the future.

\begin{appendix}\label{app}
\section*{Appendix}
\textit{Prior distributions}.
Prior distributions for the spline model parameters are specified as follows:
\begin{eqnarray*}
\exp(\lambda_{c,0}) &\sim& U(1,1000),
\\
\lambda_{c,1}/I &\sim& U(-0.25, 0.2),
\\
\chi&\sim& N(-3, 10),
\\
\varphi&\sim& U(0, 5),
\end{eqnarray*}
where $\exp(\lambda_{c,0})$ represents the level of U5MR in the
approximate midyear of the observation period,
and $\lambda_{c,1}/I$ is approximately the average ARR over the
observation period ($I$ is the interval length between knots).

Diffuse prior distributions were assigned to all data model
parameters, with the exception of the mean bias $\mu_{0,d}$ for the DHS
Direct series, which has an informative prior distribution:
\begin{eqnarray*}
\mu_{0,d} &\sim& N\bigl(M_{0,d}, S_{0,d}^2
\bigr),
\\
\mu_{1,d} &\sim& N\bigl(M_{1,d}, S_{1,d}^2
\bigr),
\\
\phi_{0,d} &\sim& U(0, 5),
\\
\phi_{1,d} &\sim& U(0, 5),
\\
\omega_{d'} &\sim& U(0,0.5),
\end{eqnarray*}
where $M_{0,d} = -0.0123$ for $d=$ DHS direct and 0 otherwise, $M_{1,d}
= 0$ for all $d$, $S_{0,d} = 0.00556$ for DHS Direct and $0.15$
otherwise, $S_{1,d} = 0.02$ for all $d$.

\textit{Logarithmic pooling approach}.
The penalized spline model-induced PPD for $\gamma_{c,K_c}^{(j)}= \Delta
\alpha_{c,K_c}^{(j)} = \alpha_{c,K_c}^{(j)} - \alpha_{c,K_c-1}^{(j)}$,
based on (\ref{eq-us}), is given by
%
\begin{equation}
\label{eq-alphaextrap} \gamma_{c,K_c}^{(j)}|\gamma_{c,K_c-1}^{(j)},
\Theta_{c,K_c}^{(j)} \sim  N \bigl(\gamma_{c,K_c-1}^{(j)},
\Theta_{c,K_c}^{(j)} \bigr),
\end{equation}
where $\Theta_{c,K_c}^{(j)} = (\sigma_{c}^2)^{(j)}$. Its density
function (leaving out superscripts to denote the posterior sample for
notational convenience) $p^*(\gamma_{c,K_c}) = f(\gamma
_{c,K_c}|\break \gamma_{c,K_c-1},  \Theta_{c,K_c})$, where $f(\Gamma|\mu, \sigma
^2)$ denotes the probability density function for a normal random
variable with mean $\mu$ and variance $\sigma^2$.\vadjust{\goodbreak}

The model-induced PPD is pooled with a (direct) global PPD for future
changes in the spline coefficients, which was based on the set of
posterior median estimates of the $\gamma_{c,k}^{(j)}$'s, $\hat{\gamma
}_{c,k}$ for $c = 1, \ldots, C$ and $k= 2,\ldots, K_c-1$ (during the
observation period for each country):
%
\begin{equation}
\label{eq-PPDglobal} p(\gamma) = f(\gamma|G,V),
\end{equation}
where $G$ and $V$ were given by the median and variance of the $\hat
{\gamma}_{c,k}$'s, respectively.

Logarithmic pooling is used to combine both density functions:
\begin{eqnarray*}
\tilde{p}(\Gamma_{c,K_c}) &\propto& p^*(\gamma _{c,K_c})^{1-w_{c,K_c}}
\cdot p(\gamma_{c,K_c})^{w_{c,K_c}}= f(\gamma _{c,K_c}|
\Gamma_{c,K_c}, \Theta_{c,K_c}),
\end{eqnarray*}
where $w_{c,K_c}$ is the country-projection-step specific logarithmic
pooling weight that determines the extent of pooling,
\[
w_{c,K_c}= \frac{W\cdot V}{W\cdot V+(1-W)\Theta_{c,K_c-1}},
\]
with overall weight $0 \leq W \leq1$ such that
\begin{eqnarray*}
\Gamma_{c,K_c} &=& W\cdot G + (1-W)\cdot\gamma_{c,K_c-1},
\\
\Theta_{c,K_c} &=& W\cdot V + (1-W)\cdot\Theta_{c,K_c-1}.
\end{eqnarray*}

For $a\geq1$, the induced PPD is defined as
\begin{eqnarray*}
p^*(\gamma_{c,K_c+a}) &=& f(\gamma_{c,K_c+a}|\Gamma_{c,K_c+a-1},
\Theta_{c,K_c+a-1}).
\end{eqnarray*}
With the global distribution from equation(\ref{eq-PPDglobal}) and
logarithmic pooling\break weights
$
w_{c,K_c+a}= \frac{W\cdot V}{W\cdot V+(1-W)\Theta_{c,K_c+a-1}}
$,
the pooled distribution for $\gamma_{c,K_c+a}$ is given by
\begin{eqnarray*}
\tilde{p}(\gamma_{c,K_c+a}) &\propto& p^*(\gamma _{c,K_c+a})^{1-w_{c,K_c+a}}
\cdot p(\gamma_{c,K_c+a})^{w_{c,K_c+a}}
\\
&=& f(\gamma_{c,K_c+a}|\Gamma_{c,K_c+a}, \Theta_{c,K_c+a}),
\\
\Gamma_{c,K_c+a} &=& W\cdot G + (1-W)\cdot\gamma_{c,K_c+a-1},
\\
\Theta_{c,K_c+a} &=& W\cdot V + (1-W)\cdot\Theta_{c,K_c+a-1}.
\end{eqnarray*}

\textit{Validation results}.
Validation results are described in Tables~\ref{tab-errors-ppd2}, \ref
{tab-errors-ppd} and \ref{tab-errors}.
\end{appendix}

\begin{table}
\caption{Validation results based on left-out observations I.
Results refer to the median (and standard deviation) of outcomes based
on 100 sets of left-out observations, where each set contains one
randomly selected observation per included country (before/including
2005, and after 2005). Included countries are given by high mortality
countries (high means U5MR of at least 40 deaths per 1000 births in
1990) without crises or HIV adjustments, with data in both training and
test set and left-out observations in the period of interest, 71 and 65
countries in total for the indicators left-out observations before and
including 2005, and left-out observations after 2005, respectively. The
outcome measures are as follows: \% of observations below and above the
90\% prediction interval based on the training set. The lowest value
for each outcome measure is bolded}
\label{tab-errors-ppd2}
\begin{tabular*}{\textwidth}{@{\extracolsep{4in minus 4in}}lcccc@{}}
\hline
& \multicolumn{4}{c@{}}{\textbf{\% of observations outside 90\% prediction
interval}}\\[-6pt]
& \multicolumn{4}{c@{}}{\hrulefill}\\
& \multicolumn{2}{c}{\textbf{Year} $\bolds{\leq 2005}$}
& \multicolumn{2}{c@{}}{\textbf{Year} $\bolds{> 2005}$} \\[-6pt]
& \multicolumn{2}{c}{\hrulefill}
& \multicolumn{2}{c@{}}{\hrulefill} \\
$\bolds{W}$ & \textbf{\% below} & \textbf{\% above} & \textbf{\% below} & \multicolumn{1}{c@{}}{\textbf{\% above}} \\
\hline
0 & 8.5 (2.6) & \textbf{7.0 (1.9)} & 9.2 (1.2) & 4.6 (1.9) \\
0.1 & \textbf{7.0 (2.6)} & \textbf{7.0 (2.0)} & \textbf{6.2 (1.2)} &
3.1 (1.3) \\
0.2 & \textbf{7.0 (2.6)} & \textbf{7.0 (2.0)} & \textbf{6.2 (1.2)} &
3.1 (1.2) \\
0.3 & \textbf{7.0 (2.5)} & \textbf{7.0 (2.0)} & \textbf{6.2 (1.2)} &
\textbf{1.5 (1.0)} \\
0.4 & \textbf{7.0 (2.5)} & \textbf{7.0 (1.9)} & \textbf{6.2 (1.3)} &
\textbf{1.5 (1.0)} \\
0.5 & \textbf{7.0 (2.4)} & \textbf{7.0 (1.8)} & \textbf{6.2 (1.5)} &
\textbf{1.5 (1.0)} \\
0.6 & \textbf{7.0 (2.5)} & \textbf{7.0 (1.7)} & \textbf{6.2 (1.5)} &
\textbf{1.5 (1.0)} \\
\hline
\end{tabular*}
\end{table}

%
\begin{table}
\caption{Validation results based on left-out observations II.
Results refer to the median (and standard deviation) of outcomes based
on 100 sets of left-out observations, where each set contains one
randomly selected observation per included country (before/including
2005, and after 2005). Included countries are given by high mortality
countries (high means U5MR of at least 40 deaths per 1000 births in
1990) without crises or HIV adjustments, with data in both training and
test set and left-out observations in the period of interest, 71 and 65
countries in total for the indicators left-out observations before and
including 2005, and left-out observations after 2005, respectively. The
outcome measures are as follows: median or mean relative error (MRE),
median or mean absolute relative error (MARE), median or mean interval
score (Score) based on the training set. The lowest value for each
outcome measure is bolded}
\label{tab-errors-ppd}
\begin{tabular*}{\textwidth}{@{\extracolsep{\fill}}ld{2.6}d{2.6}d{2.6}d{2.6}c@{}}
\hline
\multicolumn{1}{@{}l}{$\bolds{W}$} & \multicolumn{1}{c}{\textbf{ME}} & \multicolumn{1}{c}{\textbf{MAE}} &
\multicolumn{1}{c}{\textbf{MRE}} & \multicolumn{1}{c}{\textbf{MARE}} & \multicolumn{1}{c@{}}{\textbf{Score}} \\
\hline
& \multicolumn{5}{c}{Year $\leq 2005$} \\
& \multicolumn{5}{c}{Median} \\
0 & -2.2\ (1.4) & 11.3\ (1.2) & -2.2\ (1.8) & 13.2\ (1.3) &
\textbf{0.64
(0.01)} \\
0.1 & -2.2\ (1.3) & 11.2\ (1.2) & -2.2\ (1.7) & 13.2\ (1.3) &
\textbf{0.64
(0.01)} \\
0.2 & -1.9\ (1.3) & 10.9\ (1.3) & -2.1\ (1.7) & 12.9\ (1.4) & 0.64\ (0.01)
\\
0.3 & -1.9\ (1.4) & 10.8\ (1.3) & -2.1\ (1.7) & 12.9\ (1.4) & 0.64\ (0.01)
\\
0.4 & -1.9\ (1.3) & 10.8\ (1.3) & -2.0\ (1.7) & 12.9\ (1.4) & 0.64\ (0.01)
\\
0.5 & -1.7\ (1.3) & 10.7\ (1.3) & -1.9\ (1.7) & 12.9\ (1.4) & 0.64\ (0.01)
\\
0.6 & \multicolumn{1}{c}{\textbf{$\bolds{-}$1.5\ (1.3)}\phantom{0}} &
\multicolumn{1}{c}{\textbf{10.6\ (1.3)}\phantom{.}} &
\multicolumn{1}{c}{\textbf{$\bolds{-}$1.9\ (1.7)}\phantom{0}}
& \multicolumn{1}{c}{\textbf{12.9\ (1.5)}\phantom{.}} & 0.64\ (0.01) \\[3pt]
& \multicolumn{5}{c}{Mean} \\
0 & -2.9\ (1.6) & 16.6\ (1.2) & -4.4\ (1.8) & 18.6\ (1.5) & 1.03\ (0.11) \\
0.1 & -2.5\ (1.5) & 16.4\ (1.2) & -4.2\ (1.7) & 18.3\ (1.4) & 1.02\ (0.11)
\\
0.2 & -2.3\ (1.5) & 16.2\ (1.2) & -4.0\ (1.7) & 18.1\ (1.4) & 1.01\ (0.10)
\\
0.3 & -2.2\ (1.4) & 16.1\ (1.2) & -3.9\ (1.7) & 17.9\ (1.4) & 1.00\ (0.10)
\\
0.4 & -2.0\ (1.4) & 16.0\ (1.2) & -3.9\ (1.6) & 17.7\ (1.4) & 0.99\ (0.10)
\\
0.5 & -1.9\ (1.4) & 15.8\ (1.1) & -3.8\ (1.6) & 17.6\ (1.3) & 0.98\ (0.10)
\\
0.6 & \multicolumn{1}{c}{\textbf{$\bolds{-}$1.9\ (1.4)}\phantom{0}} &
\multicolumn{1}{c}{\textbf{15.6\ (1.1)}\phantom{.}} &
\multicolumn{1}{c}{\textbf{$\bolds{-}$3.7\ (1.6)}\phantom{0}}
& \multicolumn{1}{c}{\textbf{17.4\ (1.3)}\phantom{.}} &
\textbf{0.98\ (0.10)} \\[6pt]
& \multicolumn{5}{c}{Year $> 2005$} \\
& \multicolumn{5}{c}{Median} \\
0 & -3.6\ (0.4) & 10.4\ (0.6) & -10.2\ (1.1) & 17.6\ (0.7) & 0.88\ (0.03) \\
0.1 & -3.6\ (0.3) & 9.1\ (1.0) & -9.6\ (0.7) & 17.9\ (0.8) & 0.91\ (0.02) \\
0.2 & -3.7\ (0.2) & 8.4\ (1.5) &
\multicolumn{1}{c}{\textbf{$\bolds{-}$8.8\ (1.2)}\phantom{0.}} & 17.3\ (0.7) &
 0.89
(0.02) \\
0.3 & \multicolumn{1}{c}{\textbf{$\bolds{-}$3.5}\ (0.2)\phantom{0}}
& 7.6\ (1.6) & -9.2\ (1.3) & \multicolumn{1}{c}{\textbf{17.1
(0.9)}\phantom{.}} & 0.88\ (0.01) \\
0.4 & -3.6\ (0.1) &
\multicolumn{1}{c}{\textbf{7.3\ (1.4)}} & -10.0\ (0.9) & 17.2\ (1.3) &
0.86\ (0.01) \\
0.5 & -3.7\ (0.1) & 7.6\ (1.2) & -10.7\ (1.1) & 17.5\ (1.5) & 0.86\ (0.01)
\\
0.6 & -3.8\ (0.2) & 7.9\ (1.0) & -10.3\ (1.0) & 18.2\ (1.3) &
\textbf{0.84
(0.01)} \\[3pt]
& \multicolumn{5}{c}{Mean} \\
0 & -8.1\ (0.5) & 18.3\ (0.4) & -15.7\ (1.7) & 30.0\ (1.6) & 1.38\ (0.09) \\
0.1 & -7.1\ (0.5) & 17.0\ (0.5) & -14.8\ (1.5) & 28.3\ (1.4) & 1.35\ (0.08)
\\
0.2 & -6.6\ (0.5) & 16.0\ (0.5) & -14.6\ (1.4) & 27.2\ (1.3) & 1.32\ (0.08)
\\
0.3 & -6.2\ (0.5) & 15.2\ (0.5) &
\multicolumn{1}{c}{\textbf{$\bolds{-}$14.6\ (1.3)}\phantom{0.0}} & 26.6\ (1.2) &
1.30\ (0.08) \\
0.4 & -6.1\ (0.5) & 14.7\ (0.5) & -14.7\ (1.3) & 26.1\ (1.2) & 1.28\ (0.08)
\\
0.5 & -6.0\ (0.5) & 14.2\ (0.5) & -14.8\ (1.2) & 25.8\ (1.1) & 1.27\ (0.08)
\\
0.6 & \multicolumn{1}{c}{\textbf{$\bolds{-}$5.9\ (0.5)}\phantom{0}} &
\multicolumn{1}{c}{\textbf{13.9\ (0.5)}\phantom{.}} & -14.8\ (1.2) &
\multicolumn{1}{c}{\textbf
{25.6\ (1.1)}\phantom{0}} &
\textbf{1.25\ (0.08)} \\
\hline
\end{tabular*}
\end{table}

%
\begin{table}
\caption{Validation results for U5MR and ARR estimates.
Results refer to high mortality countries (high means U5MR of at least
40 deaths per 1000 births in 1990) without crises or HIV adjustments,
with data in both training and test set, 78 countries in total.
Median and mean outcome measures are reported for the U5MR in 2000 and
2005, and the annual rate of reduction (ARR) from 1990 to 2005. Outcome
measures are given by the following: median/mean relative error (MRE)
and median/mean absolute relative error (MARE) for the U5MR, median or
mean error (ME) and median/mean absolute error (MAE) for the ARR, and
median/mean interval score (Score) as well as~\% of countries below and
above the 90\% uncertainty intervals based on the training set. The
lowest value for each outcome measure is bolded}\label{tab-errors}
{\fontsize{8}{10}{\selectfont{
\begin{tabular*}{\textwidth}{@{\extracolsep{\fill}}lcccccccc@{}}
\hline
& \multicolumn{8}{c@{}}{\textbf{U5MR 2000}}\\[-6pt]
& \multicolumn{8}{c@{}}{\hrulefill}\\
&  &  & &&&& \multicolumn{2}{c@{}}{\textbf{\% of countries}} \\
& \multicolumn{3}{c}{\textbf{Median}} & \multicolumn{3}{c}{\textbf{Mean}} &
\multicolumn{2}{c@{}}{\textbf{outside 90\% UI}} \\[-6pt]
& \multicolumn{3}{c}{\hrulefill} & \multicolumn{3}{c}{\hrulefill} &
\multicolumn{2}{c@{}}{\hrulefill} \\
$\bolds{W}$ & \textbf{MRE} & \textbf{MARE} & \textbf{Score} & \textbf{MRE} & \textbf{MARE} & \textbf{Score} & \textbf{\% Below} & \textbf{\% Above} \\
\hline
0 & $-$2.4 & 4.5 & 0.30 & $-$4.8 & 9.9 & 0.56 & \textbf{3.8} & \textbf
{5.1} \\
0.1 & $-$2.4 & 4.5 & 0.30 & $-$4.6 & 9.7 & 0.56 & \textbf{3.8} & \textbf
{5.1} \\
0.2 & $-$2.4 & 4.5 & 0.30 & $-$4.5 & 9.6 & 0.55 & \textbf{3.8} & \textbf
{5.1} \\
0.3 & \textbf{$\bolds{-}$2.4} & 4.5 & 0.29 & $-$4.3 & 9.3 & 0.55 & \textbf{3.8} &
\textbf{5.1} \\
0.4 & $-$2.5 & 4.5 & 0.29 & $-$4.2 & 9.2 & 0.54 & \textbf{3.8} & \textbf
{5.1} \\
0.5 & $-$2.4 & \textbf{4.4} & 0.29 & $-$4.1 & 9.0 & 0.53 & \textbf{3.8} &
\textbf{5.1} \\
0.6 & $-$2.5 & 4.5 & \textbf{0.29} & \textbf{$\bolds{-}$4.0} & \textbf{8.8} &
\textbf{0.52} & \textbf{3.8} & \textbf{5.1} \\
\hline
\end{tabular*}}}}
{\fontsize{8}{10}{\selectfont{
\begin{tabular*}{\textwidth}{@{\extracolsep{\fill}}lcccccccc@{}}
\hline
& \multicolumn{8}{c@{}}{\textbf{U5MR 2005}}\\[-6pt]
& \multicolumn{8}{c@{}}{\hrulefill}\\
&  &  & &&&& \multicolumn{2}{c@{}}{\textbf{\% of countries}} \\
& \multicolumn{3}{c}{\textbf{Median}} & \multicolumn{3}{c}{\textbf{Mean}} &
\multicolumn{2}{c@{}}{\textbf{outside 90\% UI}} \\[-6pt]
& \multicolumn{3}{c}{\hrulefill} & \multicolumn{3}{c}{\hrulefill} &
\multicolumn{2}{c@{}}{\hrulefill} \\
$\bolds{W}$ & \textbf{MRE} & \textbf{MARE} & \textbf{Score} & \textbf{MRE} & \textbf{MARE} & \textbf{Score} & \textbf{\% Below} & \textbf{\% Above} \\
\hline
0 & $-$5.0 & 10.4 & {0.51} & $-$11.0 & 18.9 & 0.95 & \textbf{6.4} & 5.1 \\
0.1 & \textbf{$\bolds{-}$4.7} & \phantom{0}9.4 & \textbf{0.35} & $-$10.2 & 17.5 & 0.92 &
\textbf{6.4} & 3.8 \\
0.2 & $-$4.8 & \phantom{0}8.9 & 0.53 & \phantom{0}$-$9.6 & 16.4 & 0.89 & \textbf{6.4} & 2.6 \\
0.3 & $-$5.3 & \phantom{0}8.1 & 0.53 & \phantom{0}$-$9.3 & 15.7 & 0.88 & 7.7 & 2.6 \\
0.4 & $-$5.3 & \phantom{0}\textbf{8.1} & 0.54 & \phantom{0}$-$9.0 & 15.2 & 0.86 & 7.7 & 2.6 \\
0.5 & $-$6.1 & \phantom{0}8.1 & 0.53 & \phantom{0}$-$8.9 & 14.7 & 0.85 & 7.7 & \textbf{1.3} \\
0.6 & $-$6.0 & \phantom{0}8.5 & 0.53 & \phantom{0}\textbf{$\bolds{-}$8.7} & \textbf{14.4} & \textbf
{0.83} & 7.7 & 2.6 \\
\hline
\end{tabular*}}}}
{\fontsize{8}{10}{\selectfont{
\begin{tabular*}{\textwidth}{@{\extracolsep{\fill}}lcccccccc@{}}
\hline
& \multicolumn{8}{c@{}}{\textbf{ARR 1990--2005}}\\[-6pt]
& \multicolumn{8}{c@{}}{\hrulefill}\\
&  &  & &&&&\multicolumn{2}{c@{}}{\textbf{\% of countries}} \\
& \multicolumn{3}{c}{\textbf{Median}} & \multicolumn{3}{c}{\textbf{Mean}} &
\multicolumn{2}{c@{}}{\textbf{outside 90\% UI}} \\[-6pt]
& \multicolumn{3}{c}{\hrulefill} & \multicolumn{3}{c}{\hrulefill} &
\multicolumn{2}{c@{}}{\hrulefill} \\
$\bolds{W}$ & \textbf{ME} & \textbf{MAE} & \textbf{Score} & \textbf{ME} & \textbf{MAE} & \textbf{Score} & \textbf{\% Below} & \textbf{\% Above} \\
\hline
0 & 0.2 & 0.7 & \textbf{3.5} & 0.3 & 1.0 & 5.7 & 5.1 & 7.7 \\
0.1 & 0.2 & 0.6 & 3.7 & 0.3 & 1.0 & 5.5 & 5.1 & \textbf{6.4} \\
0.2 & 0.2 & 0.5 & 3.7 & 0.3 & 0.9 & 5.3 & 3.8 & 7.7 \\
0.3 & \textbf{0.1} & 0.6 & 3.7 & 0.3 & 0.9 & 5.2 & 3.8 & 7.7 \\
0.4 & 0.1 & 0.5 & 3.8 & 0.3 & 0.8 & 5.1 & \textbf{2.6} & 9.0 \\
0.5 & 0.2 & 0.5 & 3.7 & \textbf{0.3} & 0.8 & 5.0 & \textbf{2.6} & 9.0
\\
0.6 & 0.2 & \textbf{0.5} & 3.7 & 0.3 & \textbf{0.8} & \textbf{5.0} &
\textbf{2.6} & 9.0 \\
\hline
\end{tabular*}}}}
\end{table}

\section*{Acknowledgments}
The authors are very grateful to all members of the (Technical Advisory
Group of the) United Nations Inter-agency Group for Child Mortality
Estimation for passionate discussions about U5MR data and preliminary
B3 estimates which have greatly improved this work. Additional thanks
to Danzhen You, Patrick Gerland, Simon Cousens, Kenneth Hill, Kirill
Andreev, Fran\c{c}ois Pelletier, Bruno Masquelier, David Nott,\vadjust{\goodbreak} Andrew
Lover, Jakub Bijak and the Associate Editor and anonymous reviewers for
specific comments and suggestions related to the B3 model and this
manuscript, and Jon Pedersen, Jing Liu, Philip Bastian and Jingxian Wu
for database construction and assistance on data issues. We also thank
the numerous survey participants and the staff involved in the
collection and publication of the data that we analyzed. The views
expressed in this paper are those of the authors and do not necessarily
reflect the views of the United Nations Children's Fund.

%


\begin{supplement}[id=suppA]
\stitle{Figure S1: Illustration of differences in estimates and
projections for all 194 countries between the unpooled
(country-specific) and~pooled B-spline model projection approach}
\slink[doi]{10.1214/14-AOAS768SUPPA} 
\sdatatype{.pdf}
\sfilename{aoas768\_suppa.pdf}
\sdescription{Coun\-try-specific graphs to illustrate the effect of the
pooling, as in Figure~\ref{fig-ppdsimu}, for all 194 countries.}
\end{supplement}

\begin{supplement}[id=suppB]
\stitle{Figure S2: U5MR data series and estimates for all 194 countries}
\slink[doi]{10.1214/14-AOAS768SUPPB} 
\sdatatype{.pdf}
\sfilename{aoas768\_suppb.pdf}
\sdescription{Country-specific graphs, as in Figures~\ref{fig-countries} and~\ref{fig-countries2}, for all
194 countries.}
\end{supplement}

%



\printaddresses
\end{document}